\documentclass[preprintnumbers,amsmath,amssymb]{revtex4}
  \usepackage{graphicx}
\begin{document}
\title{Non-Clausius heat transfer: the example of  harmonic chain
  with an impurity}
\author{Alex V. Plyukhin}
\email{aplyukhin@anselm.edu}
 \affiliation{ Department of Mathematics,
Saint Anselm College, Manchester, New Hampshire 03102, USA 
}

\date{\today}% It is always \today, today,

\begin{abstract}
  Motivated by recent discussion about the possibility of non-Clausius
  (from cold to hot) heat flow, 
we revisit the familiar model of an impurity atom of mass $M$ embedded 
in an otherwise uniform one-dimensional harmonic lattice of host atoms of
mass $m$. 
Assuming that the initial distributions for the impurity 
and the rest of the lattice are uncorrelated and have canonical forms
with given temperatures, we  show that the average kinetic energy of the
impurity may increase
with time even if its initial temperature is higher than or equal to that
of the lattice.
Such an increase is only temporary in  uniform lattices and in  lattices
with a heavy impurity ($M\ge m$),
but may be permanent in lattices with a localized vibrational mode
generated by  a  light impurity ($M<m$). 
Thus the model shows  a non-Clausius  spontaneous heat transfer directed
from a colder lattice  to a hotter impurity.

\end{abstract}
%\pacs{05.40.-a, 66.10.C-, 82.70.Dd}

\maketitle

\section{Introduction}
According to the  Clausius formulation of the second law of thermodynamics,
heat cannot spontaneously flow from a colder system to a hotter one. 
It is often tacitly implied that 
this statement concerns only macroscopic systems.
The question of whether the statement holds at the  microscale 
has received some attention in recent years, in particular
as a part of a more general renewed debate  about
the statistical definition of entropy
(Boltzmann's versus Gibbs')
and the consistency of 
the negative temperature concept
~\cite{Gross,Dunkel,Hilbert1,Hilbert2,Campisi,Swendsen,Frenkel,Puglisi}.  

For isolated systems described by the microcanonical  ensemble,
a primary state variable
is not temperature but internal energy. On this premise,
it was argued that the sign of
the temperature difference alone does not necessarily determine the direction of
heat flow when two initially isolated systems are brought into contact;
it is thus
anticipated that under certain conditions heat can flow from
cold to hot~\cite{Gross,Hilbert1,Hilbert2,Hou}.
We shall call such anomalously directed heat transfer non-Clausius.
Other authors have disputed the possibility of non-Clausius heat
flow and criticized
the thermodynamic arguments in support of it; see~\cite{Swendsen}
and references therein.

Motivated by this discussion, in this paper we revisit the familiar model
of an impurity atom, or isotope,  embedded in a one-dimensional
harmonic lattice. The model is entirely dynamical,
except that the initial conditions for the impurity and the rest of the
lattice are given by the canonical distributions with given
(in general, different) temperatures.
We show that the initially hotter isotope  
may get energy from the colder lattice and, in the case of a light isotope,
permanently keep that energy.
Thus the terse version of the Clausius statement (heat cannot spontaneously
flow from cold to hot) 
may be  violated at the level of microscopic dynamics.
On the other hand, using a setting  discussed in the paper, one apparently
cannot design a cyclic process
to transfer heat from a cold system to a hot one in a systematic way;
therefore the second law is not violated.

Technically, our discussion is based on a generalized Langevin equation
for the impurity atom.
For the setting when the impurity and lattice are statistically uncorrelated and have different initial
temperatures, the Langevin
equation involves an additional force known as the initial slip, which
depends on the impurity's initial condition.
The initial slip often produces only transient, though possibly
long-lived,  effects~\cite{Hynes1,Hynes2,Bez,Hanggi}. On the other hand,
in  an isolated
lattice with localized  vibrational
modes~\cite{Montroll,Takeno,Kashiwamura,Rubin,Mazur,Takeno2,MM}
the initial slip is expected to be essential at all times. 
We demonstrate that 
the initial slip may be  responsible for a non-Clausius heat transfer.
The properties of the latter
strongly depend on whether the impurity generates a  localized vibrational
mode and correlate
with the ergodic properties of the model.

\section{Model}
Consider a one-dimensional lattice of $2N+1$ atoms, labeled from $-N$ to $N$,
connected by identical linear springs with the force constant $k$, see Fig. 1.
All atoms except the central one have the same mass $m$,
the central atom is an isotope with the mass $M$.
The key parameter is the mass ratio 
\begin{eqnarray}
\alpha=m/M,
\end{eqnarray}
which determines the dynamical and ergodic properties of the model. 
We shall assume that the lattice is infinitely large, $N\to\infty$,
in which case a specific choice of  boundary conditions is immaterial. 
To be specific, we shall assume that the terminal atoms are attached 
to infinitely heavy walls by the springs with the same force constant $k$ as
for the bulk of the lattice, see Fig. 1. 
The model is well familiar, but it is often discussed in the statistical
mechanics literature under the assumption
that the whole lattice is initially in thermal equilibrium, or that the uniform 
part of the lattice is initially equilibrated in the field of the
isotope fixed at a given initial position.
Here we wish to adopt  the initial condition of different kind, 
namely when the isotope and the rest of the lattice are initially equilibrated 
independently at different temperatures.

Let us write the Hamiltonian of the lattice as a sum of three terms
\begin{eqnarray}
H=H_s+H_b+H_c.
\end{eqnarray}
The term $H_s$ involves only the coordinate $Q$ and momentum
$P$ of the isotope, 
\begin{eqnarray}
H_s=\frac{P^2}{2M}+k\, Q^2.
\label{H_s}
\end{eqnarray}
We shall refer to the isotope as the system (of interest), using the
terms "isotope" and "system"  interchangeably.
The term $H_b$ involves  coordinates $\{q_i\}$ and momenta $\{p_i\}$
of all other atoms of the lattice,
which we shall refer to as the bath.
We write the Hamiltonian of the bath  as a sum of two parts corresponding
to the right and left parts of the lattice
\begin{eqnarray}
H_b&=&H_r+H_l,
\label{H_b}
\\
H_r&=&\sum_{i=1}^N \frac{p_i^2}{2m}+
\frac{k}{2}\left\{
q_1^2+(q_2-q_1)^2+\cdots+(q_N-q_{N-1})^2+q_N^2\right\},
\label{H_r}\\
H_l&=&\sum_{i=-1}^{-N} \frac{p_i^2}{2m}+
\frac{k}{2}\left\{
q_{-1}^2+(q_{-2}-q_{-1})^2+\cdots+(q_{-N}-q_{-N+1})^2+q_{-N}^2\right\}.
\label{H_l}
\end{eqnarray}
The term $H_c$ describes bilinear coupling of the system and the bath,
\begin{eqnarray}
H_c=-k\,(q_1+q_{-1})\,Q.
\label{H_c}
\end{eqnarray}
As usual, the coordinates $\{q_i\}$ and $Q$ are identified with
displacements of atoms from their mechanical equilibrium positions.

\begin{figure}[t]
\includegraphics[height=2.7cm]{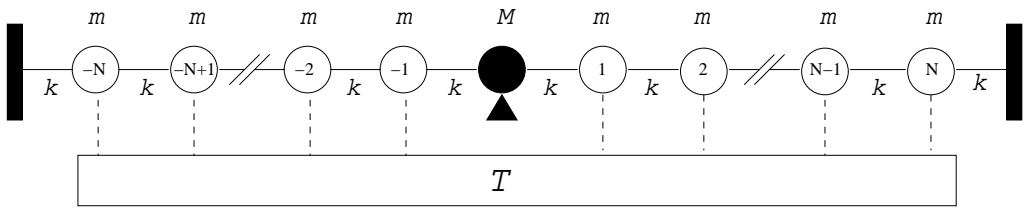}
\caption{ The lattice at times $t<0$:
  The central isotope (black circle) is fixed in the mechanical
  equilibrium position,
as denoted by the symbol $\blacktriangle$;
other atoms (white circles), referred to as the bath, are coupled
to an external thermal reservoir (depicted as a rectangle below the lattice)
of temperature $T$. 
At $t=0$ the connection to the external reservoir and the constraint
$\blacktriangle$ are removed,
and the lattice begins to evolve as an isolated system.  Initial conditions
for the bath atoms are given by the canonical distribution (\ref{rho_b}) with temperature $T$.
The initial condition for the isotope is chosen to be described
by the canonical distribution (\ref{rho_s}) with the effective
temperature $T_0$.
}
\end{figure}

The initial condition is defined as follows. 
We assume that at times $t<0$ the isotope, or the system,
is fixed in its mechanical
equilibrium position $Q=0$,  while  the rest of the lattice  (the bath)  
is in thermal equilibrium 
with an external thermal reservoir with temperatures $T$, see Fig. 1.
The state of   the lattice at $t<0$ 
can be characterized as  a state of constrained equilibrium
to emphasize  that 
an external constraint (denoted in Fig. 1 by the symbol $\blacktriangle$)
is applied to fix the initial position of the system.
At $t=0$ the contact with the external thermal reservoir and
the constraint fixing  the system's position are removed, and  
for $t>0$ the lattice evolves  as an isolated one. 

The initial distribution for the bath $\rho_b$ 
is determined by the previous contact
with the external reservoir with temperature $T$ 
and therefore has the canonical form,
\begin{eqnarray}
\rho_b=c\,e^{-H_b/T}.
\label{rho_b}
\end{eqnarray}
Here and throughout the paper we express temperature in energy units, 
i.e. multiplied by Boltzmann's constant $k_B$. 
The initial distribution for the  system $\rho_s$
we chose to be of the canonical form
\begin{eqnarray}
\rho_s=c\,e^{-H_s/T_0}
\label{rho_s}
\end{eqnarray}
with Hamiltonian $H_s$ given by (\ref{H_s}) and the parameter $T_0$ playing the 
role of the isotope's initial temperature.
Such initial condition 
can be characterized as a sudden preparation~\cite{Bez}
in the sense that the bath has no time to adjust to initial
parameters of the system.
One may interpret the initial distribution (\ref{rho_s}) as follows:
Immediately after it is released 
at $t=0$, the isotope is harmonically trapped in the potential $k\,Q^2$
and reaches thermal equilibrium with an external thermal
reservoir with temperature $T_0$ on a timescale much shorter 
than any other characteristic times of the problem.
Note that Hamiltonian $H_s$ in (\ref{rho_s}) depends on the coordinate and momentum 
of the system only, and does not involve interaction with the bath. This reflects 
our assumption that the system and bath are initially thermalized independently.

Already at this stage of the discussion, one may anticipate  that the model
may show properties which are unusual from the point of view of macroscopic
thermostatistics.
Even if the initial temperatures of the system and the bath are the same, 
$T_0=T$, 
the overall lattice at $t=0$ is not  in thermal equilibrium.  Indeed,
the initial distribution  $\rho_s\rho_b\sim e^{-(H_s+H_b)/T}$ 
does not involve the coupling Hamiltonian $H_c$, and therefore 
is not the equilibrium distribution 
$\rho_e\sim e^{-H/T}=e^{-(H_s+H_b+H_c)/T}$ for the lattice.
This feature is not a surprise, 
considering that in macroscopic thermostatistics the interaction energy of
the system and the bath is assumed to be negligible compared to the internal
energy of the system. 
On the other hand, for microscopic systems
strongly coupled to the environment the interaction energy in not 
negligible, and one
may anticipate a transient heat flow between the system and bath even 
if their temperatures are the same.

\section{Langevin equation}
The dynamics of the model 
described in the previous section can be analyzed in many ways; 
here we use the method based on a microscopically derived  Langevin equation
for the isotope.
Integrating equation of motion for the bath atoms, substituting the results 
into the equation of motion for the isotope,
and taking the limit of the infinite lattice $N\to\infty$, one derives for
the isotope's momentum the following generalized Langevin equation:
\begin{eqnarray}
\dot P(t)= -\int_0^t K(t-\tau)\,P(\tau)\,d\tau+\xi(t)-M\,Q(0)\,K(t). 
\label{GLE}
\end{eqnarray}
This equation has been the subject of many works, see e.g.
\cite{Bez,Hanggi,Weiss}.
In order to make the paper self-contained we provide the
derivation in Appendix A.
In Eq. (\ref{GLE}), the term  $\xi(t)$ is a stochastic force which for the
present model is available as an explicit linear function of initial coordinates
and momenta of the bath $\{q_i,p_i\}$,
see Eq. (\ref{stochastic_force_chain}).
One can show that $\xi(t)$ is
a zero-centered and stationary process  
related to the memory kernel $K(t)$ by the standard
fluctuation-dissipation relation,
\begin{eqnarray}
\langle\xi(t)\rangle=0, \qquad \langle \xi(t)\,\xi(t')\rangle=M\,T\,K(t-t'). 
\label{fdt}
\end{eqnarray}
Here the average is taken over 
initial coordinates and momenta of the bath 
$\{q_i,p_i\}$
with the distribution $\rho_b$ given by Eq. (\ref{rho_b}).
Let us note that although properties (\ref{fdt}) 
for $\xi(t)$ 
are standard, they should not be taken for granted.
In general, when the bath  initially is not in equilibrium with the system,
as in
our model, the stochastic  force  $\xi(t)$  
may have more complicated properties~\cite{Maes}.

The memory kernel $K(t)$ in Eq. (\ref{GLE})
for the infinite lattice is known to have the form
\begin{eqnarray}
  K(t)=\frac{\alpha\,\omega_0}{t}\,J_1(\omega_0 t)=
  \frac{\alpha\,\omega_0^2}{2}\, \Bigl\{
J_0(\omega_0 t)+J_2(\omega_0 t)
\Bigr\},
\label{kernel}
\end{eqnarray}
where $J_n(x)$ are Bessel functions, $\alpha=m/M$
and $\omega_0=2\sqrt{k/m}$, see Appendix A.
We shall also need the Laplace transform $\tilde K(s)=
\int_0^\infty e^{-st}\,K(t)\,dt$
of kernel (\ref{kernel}), which reads
\begin{eqnarray}
\tilde K(s)=\frac{\alpha\,\omega_0^2}{s+\sqrt{s^2+\omega_0^2}}=\alpha\,\left(
-s+\sqrt{s^2+\omega_0^2}\right).
\label{kernel_Laplace}
\end{eqnarray}

A distinctive feature of the Langevin equation (\ref{GLE})
is the 
force 
$-M\,Q(0)\, K(t)$, known as the initial slip~\cite{Hynes1,Hynes2,Bez,Hanggi},
which is a linear function of the isotope's initial displacement $Q(0)$.
%The presence of such a force, not necessarily in the form as in Eq. (\ref{GLE}),
%is a common feature of the Langevin dynamics when the system and the bath are
%initially not in equilibrium. 
The initial slip vanishes for the initial condition $Q(0)=0$, but
it does not appear in the Langevin equation  also for special
initial distributions with $Q(0)\ne 0$. In particular, if initial values of the bath variables
are drawn from the distribution
\begin{eqnarray}
  \rho_c=c\,e^{-[H_b+H_c]/T}
\label{rho_c}
\end{eqnarray}
then the initial slip has to be absorbed in the  
stochastic force $\xi(t)$ in order to make the latter
zero-centered~\cite{Weiss,Lindenberg}.  Distribution (\ref{rho_c}), 
which is often exploited in microscopical models of open systems 
at strong coupling (see, e.g., Refs.~\cite{MO,Seifert}), 
implies that the bath at $t=0$ is in equilibrium with the system fixed at a given position. 
This assumes that the bath variables are fast and  quickly adjust (reach equilibrium with) 
the slowly moving system. In this paper we keep in mind a quite different 
physical situation when the system and bath at $t<0$ do not interact and are 
individually equilibrated to canonical states  with different temperatures.
Respectively, instead of using distribution (\ref{rho_c}) we assume that 
initial states of the bath and system are not correlated and described 
by canonical distributions $\rho_b$ and $\rho_s$, given by Eqs. (\ref{rho_b}) 
and (\ref{rho_s}). In that case the initial slip in the Langevin equation (\ref{GLE}) 
does not vanish and will be shown to play an important role.

\section{Thermalization and non-thermalization}

Solving the Langevin equation (\ref{GLE}) with the Laplace transform
method one gets
\begin{eqnarray}
P(t)=P(0)\,R(t)+\int_0^t R(\tau)\,\xi(t-\tau)\, d\tau+M\, Q(0)\,\dot{R}(t),
\label{sol1}
\end{eqnarray}
where the dimensionless function $R(t)$ has the Laplace transform
\begin{eqnarray}
\tilde R(s)=\frac{1}{s+\tilde K(s)}.
\label{resolvent1}
\end{eqnarray}
In the time domain $R(t)$ satisfies  the following equation
and initial condition:
\begin{eqnarray}
\dot R(t)=-\int_0^t K(t-\tau)\,R(\tau)\,d\tau, \qquad R(0)=1.
\label{resolvent2}
\end{eqnarray}

We shall call the function $R(t)$ the resolvent.
As follows from the above relations, the resolvent $R(t)$ is a solution of
the generalized Langevin equation (\ref{GLE}) for
for the special initial condition when $P(0)=1$,
$Q(0)=0$, and initial displacements and momenta of all atoms of the bath
are zero, $q_i(0)=p_i(0)=0$ for  $\forall i$. 
Indeed, as shown in Appendix A,
the fluctuating force $\xi(t)$ is a linear function of initial bath variables.
Then for the above mentioned initial conditions
$\xi(t)=0$ for any $t$, and Eq. (\ref{sol1}) gives $P(t)=R(t)$.

The resolvent $R(t)$ is also equal to a more familiar character of  statistical
mechanics, 
namely 
the normalized equilibrium
correlation function 
\begin{eqnarray}
C(t)=\frac{\langle P(0) P(t)\rangle_e}{\langle P^2\rangle_e}.
\label{C}
\end{eqnarray}
Here the average $\langle\cdots\rangle_e$ is taken
over initial coordinates and momenta of the overall lattice with the
canonical equilibrium distribution $\rho_e=c\,e^{-H/T}$.
The equality $C(t)=R(t)$ can be
verified by 
constructing from (\ref{sol1}) the expression for $C(t)$.
Alternatively, one can notice 
that when the system and the bath 
are in equilibrium, 
the  Langevin equation for the system  has the standard form
with no initial slip,
\begin{eqnarray}
\dot P(t)= -\int_0^t K(t-\tau)\,P(\tau)\,d\tau+\xi(t),
\label{GLE2}
\end{eqnarray}
where $\xi(t)$ is zero-centered, $\langle \xi(t)\rangle_e=0$, and
uncorrelated with $P(0)$. 
Multiplying this equation by $P(0)$, taking the average and normalizing,
one finds that  $C(t)$ satisfies the initial value problem
\begin{eqnarray}
\dot C(t)=-\int_0^t K(t-\tau)\,C(\tau)\,d\tau, \qquad C(0)=1,
\end{eqnarray}
which is the same as Eq. (\ref{resolvent2}) 
as for the resolvent. 
The two functions therefore are equal, $R(t)=C(t)$.
An explicit expression for the equilibrium  correlation function $C(t)$
for the present model is well-known~\cite{Rubin}; we shall exploit that
result identifying $C(t)$ with the resolvent $R(t)$.

Within the Langevin approach, it is natural to describe the system's dynamics
by evaluating the moments $\langle P^n(t)\rangle$. We shall focus on the first
two moments, $\langle P(t)\rangle$ and $\langle P^2(t)\rangle$, taking
the average $\langle \cdots\rangle$ over 
the initial bath variables $\{q_i(0),p_i(0)\}$ with the canonical
distribution $\rho_b$ given by Eq. (\ref{rho_b}). 

The expression for the first moment follows immediately  from Eq. (\ref{sol1}),
\begin{eqnarray}
\langle P(t)\rangle=P(0)R(t)+M\,Q(0) \dot R(t).
\label{moment1}
\end{eqnarray}
Squaring Eq. (\ref{sol1}) and taking the average, one gets for the second
moment the expression
\begin{eqnarray}
\langle P^2(t)\rangle=P^2(0)\,R^2(t)+
\int_0^t \!\!d\tau\!\!\int_0^t \! d\tau'\,R(\tau)\,R(\tau')\,
\langle \xi(\tau)\xi(\tau')\rangle+
[M\,Q(0)\, \dot R(t)]^2+M\,P(0)\,Q(0)\, \frac{d}{dt}\,R^2(t). 
\label{p2}
\end{eqnarray}
Here we took advantage of the stationarity of the stochastic force,
$\langle \xi(t-\tau)\,\xi(t-\tau')\rangle=\langle \xi(\tau)\xi(\tau')\rangle$. 
The double  integral over the square $[0,t]\times [0,t]$ can be written as two
times the integral over a triangle,
\begin{eqnarray}
2\int_0^t d\tau \,R(\tau) \int_0^{\tau} d\tau'\,R(\tau')\,
\langle \xi(\tau)\xi(\tau')\rangle.
\label{integral1}
\end{eqnarray}
Using the fluctuation-dissipation relation (\ref{fdt})
and Eq. (\ref{resolvent2}),
this can be further worked out as follows:
\begin{eqnarray}
2M\, T\int_0^t d\tau\, R(\tau) \int_0^{\tau} d\tau'\,K(\tau-\tau')\,R(\tau')
=-2M\, T\int_0^t d\tau\, R(\tau)\, \dot R(\tau)
=M\, T \,\left[1-R^2(t)\right].
\label{integral2}
\end{eqnarray}
Then expression (\ref{p2}) 
takes the form
\begin{eqnarray}
  \langle P^2(t)\rangle=P^2(0)\,R^2(t)+M\,T\,[1-R^2(t)]+
          [M\, Q(0)\,\dot R(t)]^2+M\,P(0)\,Q(0)\, \frac{d}{dt}\,R^2(t).
\label{moment2}
\end{eqnarray}

As follows from (\ref{moment1}) and (\ref{moment2}), the ergodic properties
of the model are determined by the asymptotic properties of the resolvent
and its derivative: if $R(t)$ and $\dot R(t)$  both vanish
in the long time limit, 
\begin{eqnarray}
  R(t)\to 0,\quad \dot R(t)\to 0, \quad \mbox{as} \quad t\to\infty,
  \label{thermalization_condition}
\end{eqnarray}
then the moments thermalize, i.e. 
evolve towards the equilibrium 
values, 
\begin{eqnarray}
\langle P(t)\rangle\to 0, \qquad
\langle P^2(t)\rangle\to M\,T,
\quad \mbox{as} \quad t\to\infty.
\label{equilibrium}
\end{eqnarray}
For the present model thermalization is known to occur for a heavy isotope
or for a uniform lattice, $M\ge m$, or $\alpha\le 1$~\cite{Rubin,Weiss, Mazur}.
Even though the  harmonic lattice is not an ergodic system, a heavy
isotope embedded in the lattice shows the ergodic behavior.  
On the other hand, for a light isotope, $M<m$, or $\alpha>1$ 
the resolvent $R(t)$ involves
a non-vanishing oscillating component known as a localized vibrational
mode~\cite{Rubin,Montroll,Takeno,Kashiwamura,Mazur}. In that latter case the 
condition (\ref{thermalization_condition}) is not satisfied, and the 
light isotope does not reach thermal  equilibrium
with the bath.

We postpone the further discussion of ergodic properties of the model and
the evaluation of the resolvent until Sec. VI.

\section{Heat transfer}
Recall that expressions (\ref{moment1})
and (\ref{moment2}) for the moments $\langle P(t)\rangle$
and $\langle P^2(t)\rangle$ involve averaging over the bath variables only. Now
 we take an additional average
 of that expressions
$\langle \cdots\rangle_s=\int dQ\, dP\, \rho_s (\cdots)$
over the initial isotope's
coordinate and momentum  $Q=Q(0)$, $P=P(0)$
with distribution $\rho_s$  given by (\ref{rho_s}), 
\begin{eqnarray}
  \langle P(t)\rangle&=&\langle P\rangle_s R(t)+M\,\langle Q\rangle_s\dot R(t),
  \nonumber\\
  \langle P^2(t)\rangle&=&\langle P^2\rangle_s\,R^2(t)+M\, T\,[1-R^2(t)]+
  M^2\, \langle Q^2\rangle_s \,[\dot R(t)]^2+M\,\langle P\,Q\rangle_s\, \frac{d}{dt}\,R^2(t).
\label{moment12}
\end{eqnarray}
With $H_s$ given by Eq. (\ref{H_s}), 
the distribution  
$\rho_s=c\,e^{-H_s/T_0}$
describes
the  equilibrium state  of a harmonic oscillator with mass $M$ and
spring constant $2k$, and therefore has the moments
\begin{eqnarray}
  \langle P\rangle_s=\langle Q\rangle_s=\langle P\,Q\rangle_s=0,\quad\langle P^2\rangle_s=
  M\,T_0,\quad
\langle Q^2\rangle_s=\frac{1}{2k}\,T_0.
\label{IC}
\end{eqnarray}
Substituting these values into 
Eq. (\ref{moment12}), we find  that the first moment vanishes,
$\langle P(t)\rangle=0$,
and the second moment takes the form 
\begin{eqnarray}
\langle P^2(t)\rangle=
M\,T_0\,R^2(t) +
M\,T[1-\,R^2(t)]+\frac{2M}{\alpha\,\omega_0^2}\,T_0\,[\dot R(t)]^2.
\label{moment2_3}
\end{eqnarray}
Here we take into account
that $M/k=4/(\alpha\,\omega_0^2)$. Thus, the  kinetic energy of
the system $E=\langle P^2\rangle/2M$, averaged over initial variables
of both the system and bath, reads
\begin{eqnarray}
E(t)=
\frac{T_0}{2}\,R^2(t) +\frac{T}{2}\,[1-\,R^2(t)]+
\frac{T_0}{\alpha\,\omega_0^2}\,[\dot R(t)]^2.
\label{E}
\end{eqnarray}
Our interest is to compare $E(t)$ with
the system's initial kinetic energy
$E(0)= T_0/2$ evaluating 
the energy change 
\begin{eqnarray}
\Delta E(t)=E(t)-E(0)=E(t)-T_0/2.
\label{D_def}
\end{eqnarray}
The sign of  $\Delta E(t)$ 
characterizes the direction of the net heat transfer during
the time interval $(0,t)$:
$\Delta E(t)>0$ suggests that 
the system absorbs heat from the bath, while 
$\Delta E(t)<0$ corresponds to the transfer of heat in the opposite direction. 
From (\ref{E}) and (\ref{D_def}) one obtains
\begin{eqnarray}
\Delta E(t)=
\frac{1}{2}\,[1-\,R^2(t)]\,(T-T_0)+\frac{1}{\alpha\,\omega_0^2}\,[\dot R(t)]^2\,T_0.
\label{D}
\end{eqnarray}
In the following sections we shall
discuss the explicit forms which the function  $\Delta E(t)$ takes for the cases
of heavy and light isotopes. However, some general features of the heat transfer
can be observed already at this stage based only on the  asymptotic properties
of the resolvent.

As was discussed in Section IV, for a heavy isotope
($\alpha\le 1$) the resolvent $R(t)$ and its derivative $\dot R(t)$
both vanish at long times. As a consequence, the dynamics of a
heavy isotope is irreversible and ergodic:  the isotope  reaches thermal
equilibrium with the lattice. From Eq. (\ref{D})  we find that for
a heavy isotope the kinetic energy change reaches the long-time
asymptotic value
\begin{eqnarray}
\Delta E(t)\to\frac{1}{2}\,(T-T_0),\qquad t\to\infty,
\label{D_limit}
\end{eqnarray}
which corresponds to a Clausius heat 
transfer from hot to cold.  
The same value we obtain for 
$\Delta E(t)$ averaged over an infinitely  long time interval,
\begin{eqnarray}
\overline{\Delta E}=\lim_{t\to\infty}\frac{1}
{t}\int_0^{t} \Delta E(\tau) \,d\tau.
\label{delta_av}
\end{eqnarray}
Thus we find that for the present model 
the ergodic dynamics also implies that 
the net heat transfer between the system and the bath on the long
time scale is Clausius.

On the other hand, one can observe from Eq. (\ref{D})  that  Clausius heat
transfer is a property that should not be taken for granted.
In general, it only holds for ergodic systems, and only on the asymptotically
long time interval.
For a non-ergodic system, like a light isotope, $R(t)$ does not vanish at
long times. In that case,  as we shall see in Section VIII,
Eq. (\ref{D}) may give a non-Clausius
heat transfer. 
Moreover, even for the ergodic dynamics of a heavy isotope Eq. (\ref{D}) 
predicts that heat transfer may be non-Clausius on a finite
time interval.  
The first term in the
right-hand side of Eq. (\ref{D}) is  proportional to the temperature difference $T-T_0$ and thus   
describes a Clausius net heat transfer.
On the other hand, the second term in Eq. (\ref{D}) 
does not involve the temperature difference, but 
depends on the initial temperature of the system only. 
Being non-negative,
the second term describes a heat flow directed from the bath 
to the system even for $T_0\ge T$.
Then the sign of the net kinetic energy balance may 
depend not only on 
the temperature difference, but also on the relative 
strength of the two terms in Eq. (\ref{D}).

Suppose $T>T_0$, i.e. the bath is hotter than the system.
For any $t>0$ in (\ref{D}) the first term is positive and the
second one is non-negative, so that $\Delta E(t)>0$. Thus, for $T>T_0$
the net heat transfer is always Clausius: the colder system gets a
positive amount of energy from  the hotter bath. 
However, since the resolvent $R(t)$ is, in general, a non-monotonic function, 
Eq. (\ref{D}) shows that on finite time intervals $\Delta E(t)$ may be negative.

When $T=T_0$ the first term in (\ref{D}) is zero, and $\Delta E(t)\ge 0$.
The heat transfer is non-Clausius:
the system receives some energy from the bath despite  the temperatures of 
the system and the bath are the same.
When the system is a heavy isotope ($\alpha\le 1$), then $\dot R(t)\to 0$
as $t\to\infty$, and according to (\ref{D}), $\Delta E(t)$ vanishes at long times.
In this case the non-Clausius heat transfer is transient and vanishes in
the long time limit.
On the other hand, if the system is a light isotope, $\alpha>1$,
then $\dot R(t)$ does not vanish but rather oscillates at long times.
In thus case, the time average of $\Delta E(t)$, defined
by (\ref{delta_av}), is positive, $\overline{\Delta E}>0$, and the non-Clausius
heat transfer is permanent.

Now suppose $T<T_0$, i.e. the isotope is initially hotter than the bath.
In this case the two terms in (\ref{D}) have different signs and, as we noted above,
the  net heat transfer may be either Clausius or non-Clausius depending
on the relative values of the first and second terms.
It is clear, however, that for a small temperature difference the first
term, at least at short times,  is smaller by the absolute value
then the second one,
so that  $\Delta E(t)>0$. This corresponds to a non-Clausius heat transfer: 
the initially hotter system receives energy from the colder bath.
Similarly to the case $T=T_0$, the system's energy gain is
only temporary for $\alpha\le 1$ but may be  permanent for $\alpha>1$.

For a further analysis and illustrations we need  explicit expressions
for the resolvent $R(t)$.

\section{Resolvent, ergodicity, and localized modes}
As discussed in the previous sections, 
the ergodic and heat transfer properties of the present model are governed by
the asymptotic long-time properties of the resolvent $R(t)$. 
As follows from Eqs. (\ref{resolvent1}) and (\ref{kernel_Laplace}),
the Laplace transform of the resolvent reads
\begin{eqnarray}
\tilde R(s)=\frac{1}{\alpha \sqrt{s^2+\omega_0^2}-(\alpha-1) s}.
\label{resolvent_Laplace}
\end{eqnarray}
Recall that here $\alpha=m/M$ is the mass ratio,
and $\omega_0=2\sqrt{k/m}$ is the maximal normal mode frequency
of the uniform lattice.
The inversion of (\ref{resolvent_Laplace}) in closed form is possible
only for a uniform lattice,
\begin{eqnarray}
R(t)=J_0(\omega_0 t), \quad
\mbox{if} \quad
\alpha=1,
\label{case1}
\end{eqnarray}
and for a heavy isotope two times heavier than atoms of the chain,
\begin{eqnarray}
R(t)=\frac{2}{\omega_0 t}\,J_1(\omega_0 t), \quad
\mbox{if}\quad
\alpha=1/2,
\label{case2}
\end{eqnarray}
where $J_n(x)$ are Bessel functions. In both cases the resolvent and
its derivative vanish at long times and,  according to
(\ref{moment2}), the isotope reaches thermal
equilibrium with the bath.
It turns out that this scenario holds also for any $\alpha\le 1$.
For that case the inversion of (\ref{resolvent_Laplace})
gives the following result:
\begin{eqnarray}
  R(t)=\varphi(t)=
  \frac{2\alpha}{\pi}\int_{0}^{\omega_0}\frac{\sqrt{\omega_0^2-y^2}\,\cos(yt)}
{(1-2\alpha)y^2+\alpha^2\omega_0^2}\,dy,
\quad
\mbox{if} \quad
\alpha\le 1.
\label{R_heavy}
\end{eqnarray}
One can verify that the function $\varphi(t)$ given by this expression
decays to zero at long times for any $\alpha$.
For $\alpha=1$ and $\alpha=1/2$ expression (\ref{R_heavy})
is reduced to functions (\ref{case1}) and (\ref{case2}), respectively.

For a light isotope $\alpha>1$,
the inversion of (\ref{resolvent_Laplace}) gives a qualitatively different result:
\begin{eqnarray}
R(t)=A(\alpha)\,\cos(\omega_* t)+\varphi(t), 
\quad
\mbox{if} \quad
\alpha> 1.
\label{sol2}
\end{eqnarray}
Here the function $\varphi(t)$ 
is still given by Eq. (\ref{R_heavy}) and vanishes
at long times. The resolvent, however, does not vanish and is
given for long times by the first term oscillating with
the frequency and amplitude
\begin{eqnarray}
\omega_*=\frac{\alpha}{\sqrt{2\alpha-1}}\,\omega_0>\omega_0, \qquad
A(\alpha)=\frac{2\alpha-2}{2\alpha-1}<1.
\label{omega*}
\end{eqnarray}
As discussed in Sec. IV, see Eq. (\ref{moment2}), for a resolvent $R(t)$ non-vanishing
at long times the isotope does not reach thermal
equilibrium with the bath. Thus the dynamics of a light isotope
embedded in a harmonic chain is
non-ergodic~\cite{Montroll,Takeno,Kashiwamura,Rubin,Mazur}.

The results (\ref{R_heavy})-(\ref{omega*})
are well known~\cite{Rubin}; in Appendix B we give their derivation,
i.e. the inversion of the  transform (\ref{resolvent_Laplace}),
in full detail. 
Note that the initial value of the function $\varphi(t)$ depends on the mass ratio, 
namely
$\varphi(0)=1$ for $\alpha\le 1$, and $\varphi(0)=1-A(\alpha)$ for $\alpha>1$. 
As a result, both expressions (\ref{R_heavy}) and (\ref{sol2}) 
for the resolvent satisfy the initial condition $R(0)=1$.

The presence of a localized vibrational mode with frequency $\omega_*$
associated with a light isotope is 
essential for our further discussion. 
When excited in an isolated lattice, such a vibration does not dissipate
by generating running waves but lasts forever
 as a localized oscillation of the isotope and its neighbours.
 The phenomenon is similar to localization of electrons in solids
 around impurity centers.
 The general condition  of normal mode  localization is that its
 frequency $\omega_*$ is outside the frequency spectrum of
 the undisturbed lattice~\cite{Montroll,Rubin,Takeno,Kashiwamura}.
 For a uniform  harmonic chain with a single impurity atom that condition
 is satisfied when the impurity is a light isotope, $\alpha>1$,
 see also~\cite{Rubin,Takeno,Kashiwamura}. On the other hand, for
 lattice systems with a more complicated composition
 the localization condition does not necessarily implies
 $\alpha>1$~\cite{Takeno2,MM,IT}. Localized modes may emerge not only due
 to defects, but also due to the lattice's  anharmonicity~\cite{Flach}.

Technically, the emergence of a localized mode can be understood as follows. 
Given the transform $\tilde R(s)$,
the resolvent $R(t)$ in the time domain is given by 
 the Bromwich integral in the complex plane,
 $R(t)=\frac{1}{2\pi i}
 \int_{\gamma-i\infty}^{\gamma+i\infty} e^{st}\, \tilde R(s)\,ds$.
 Following the usual routine,  one considers a corresponding auxiliary
 integral $I(t)=\frac{1}{2\pi i} \int_\Gamma e^{st}\, \tilde R(s)\,ds$
over a properly closed contour $\Gamma$, see the right side of Fig. 4.  
The integral $I(t)$
can be evaluated using  Cauchy's residue  theorem. For a light isotope,
$\alpha>1$,  the function $\tilde R(s)$ given by (\ref{resolvent_Laplace})
has simple poles at $\pm i \omega_*$,
with $\omega_*$ given by (\ref{omega*}),
located on the imaginary axis.
Then the integral $I(t)$ 
has components given by the residues 
${\rm Res} [e^{st}\tilde R(s), i\omega_*]$
and
${\rm Res} [e^{st}\tilde R(s), -i\omega_*]$, which oscillate with 
frequency $\omega_*$. Together these components form the oscillating
term in expression (\ref{sol2}) for resolvent
$R(t)$. Appendix B provides further details.

It may appear from the above reasoning 
that localization may occur also for a heavy isotope with $1/2<\alpha\le 1$,
since in that case expression (\ref{omega*}) for $\omega_*$ is
real and therefore 
$\tilde R(s)$ has the poles $\pm i\omega_*$ on the imaginary axis, as for a
light isotope.
However, one has to take into account that the function $\tilde R(s)$
given by (\ref{resolvent_Laplace}) has two branches, and only one of them
is physically meaningful, satisfying the initial condition $R(0)=1$.
As shown in Appendix B, for 
$\alpha\le 1$ 
the physical branch of  $\tilde R(s)$ has no poles, and the inversion
gives the result (\ref{R_heavy}).
Thus, for a heavy isotope and a uniform lattice , $\alpha\le 1$,
there is no localization.

Now equipped with explicit expressions for the resolvent,
we can return to the discussion of the energy exchange
between the isotope and the bath.

\section{Uniform lattice and lattice with a heavy isotope}

Consider first the case  $\alpha=1$ when the system is just
a tagged atom in a uniform lattice. The resolvent is given by (\ref{case1}),
$R(t)=J_0(\omega_0 t)$, and 
expression (\ref{D}) for 
the relative energy change function $\Delta(t)$ takes the form
\begin{eqnarray}
\Delta E(t)=\frac{1}{2}\,[1-J_0^2(\omega_0 t)]\,(T-T_0)
+J_1^2(\omega_0 t)\,T_0
\label{D1}
\end{eqnarray}
As was noted above, the long time limit, as well as the time
average $\overline{\Delta E}$,
of this expression,
\begin{eqnarray}
\lim_{t\to\infty}\Delta E(t)=\overline{\Delta E}=\frac{1}{2}\,(T-T_0),
\label{D_limit2}
\end{eqnarray}
corresponds to Clausius heat transfer for any values of both temperatures.  
Moreover, while the function $\Delta E(t)$ is not monotonic,
for $T>T_0$ (the bath is initially hotter than the isotope), it is 
positive for any time $t>0$,
see Fig. 2(a).

For $T=T_0$, the energy change function (\ref{D1}) is non-negative for any $t$,
see Fig. 2(b), which 
corresponds to a non-Clausius heat flow 
from the bath to the system, despite no temperatures difference is imposed.
At long times, however, $\Delta E(t)$ goes to zero, and so does the heat flow. 

\begin{figure}[t]
\includegraphics[height=8.5cm]{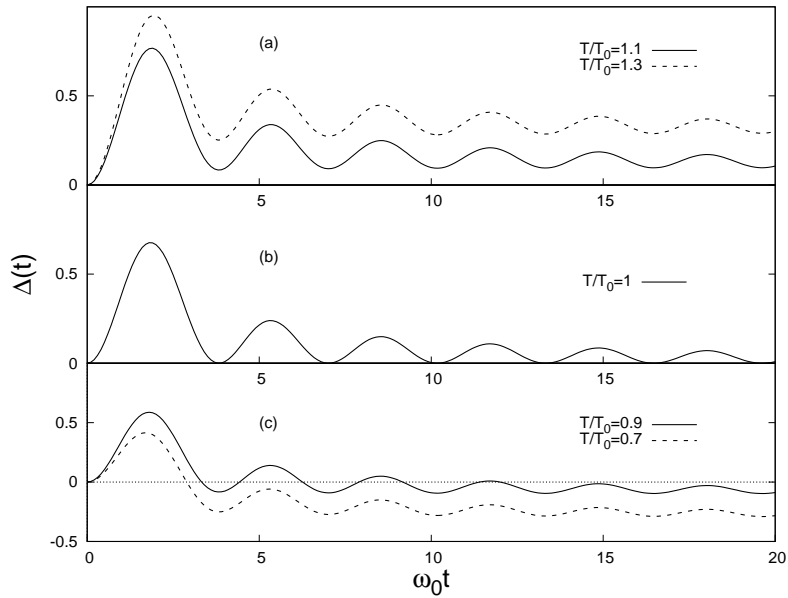}
\caption{ 
The average kinetic energy change $\Delta E(t)=E(t)-E(0)$, in units $E(0)=T_0/2$, 
as a function of time for a tagged atom in a uniform lattice, $\alpha=1$, 
according to Eq. (\ref{D1}),
for different initial temperatures of the atom ($T_0$) and the lattice ($T$).   
}
\end{figure}

\begin{figure}[t]
\includegraphics[height=8.5cm]{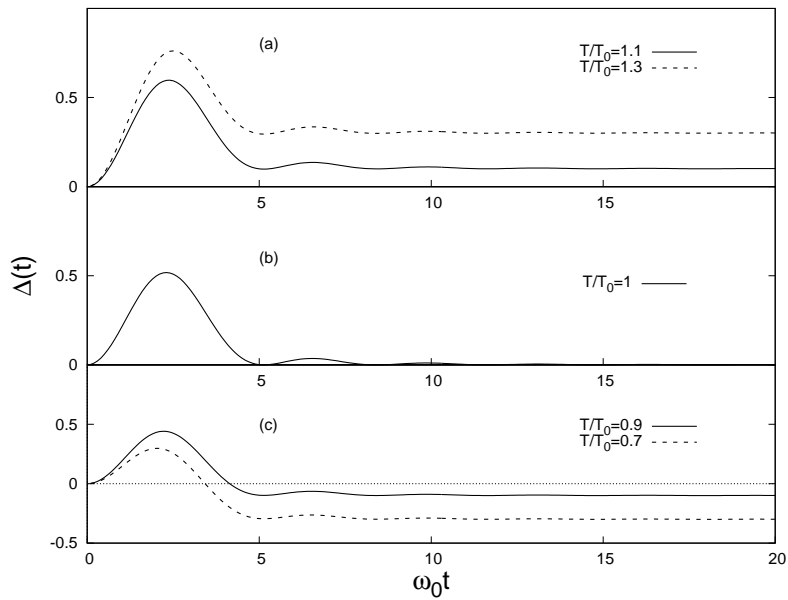}
\caption{ Same as Fig. 2, but for a heavy
  isotope with $\alpha=1/2$, according to Eq. (\ref{D2}).
}
\end{figure}

For $T<T_0$ (the bath is initially colder than the system),
the  function $\Delta E(t)$ is positive on the initial and possibly
(if the temperature difference is sufficiently small) on later
finite time intervals, see Fig. 2(c).
During those intervals the heat transfer is non-Clausius:
$\Delta E(t)>0$ for $T<T_0$ means that the initially hotter
system obtains energy from the  colder bath. 
However, at long times $\Delta E(t)$ becomes negative,
so that the heat flow turns the direction and 
becomes Clausius.

For a lattice with  a heavy isotope $\alpha<1$ the situation
is qualitatively similar to that for a uniform lattice, $\alpha=1$.
For a  hotter lattice $T>T_0$ the heat transfer is Clausius at all times.
For a hotter isotope, $T<T_0$, or equal temperatures $T=T_0$, 
the heat transfer is initially    
 non-Clausius,  but becomes Clausius on a longer time scale.
 Fig. 3 illustrates this behavior for a heavy isotope  with $\alpha=1/2$.
 In that case  the resolvent is available in closed form,
 $R(t)=2J_1(\omega_0 t)/(\omega_0 t)$, and  the energy change (\ref{D})
 takes the form
\begin{eqnarray}
  \Delta E(t)=\frac{1}{2}\,\left\{1-\frac{4\,J_1^2(\omega_0 t)}{(\omega_0 t)^2}\right\}
  \,(T-T_0)+\frac{8}{(\omega_0 t)^2}
\,J_2^2(\omega_0 t)\,T_0.
\label{D2}
\end{eqnarray}
Comparing Fig. 2 and Fig. 3, one observes that the plots for a uniform
lattice and a heavy isotope are 
qualitatively 
similar, but in the latter  case the oscillatory tail of $\Delta E(t)$
decays faster.

For a heavy isotope with other values of the mass ratio $\alpha<1$ the 
resolvent $R(t)$
and energy change $\Delta E(t)$ can be evaluated numerically 
using for the former Eq. (\ref{R_heavy}). In all cases we observed
the same behavior as for $\alpha=1$ and $\alpha=1/2$:
the heat transfer may be non-Clausius on a short time scale 
but is back to normal at longer times.

\section{Lattice with a light isotope}

As discussed in Sec. VI, when the system is a light isotope, $\alpha>1$,
the resolvent has a non-decaying oscillatory component
due to the formation of a localized mode,
$R(t)=A\cos(\omega_* t)+\varphi(t)$. At long times the function $\varphi(t)$
vanishes,
and $R(t)\approx A\cos(\omega_* t)$. Respectively, at long times 
the energy change function (\ref{D}) takes the form
\begin{eqnarray}
\Delta E(t)=\frac{1}{2}\,\left[1-A^2 \cos^2(\omega_* t)\right]
\,(T-T_0)+\frac{1}{\alpha}\,
\left(\frac{\omega_*}{\omega_0}\right)^2 A^2\sin^2(\omega_0 t)\,T_0
\end{eqnarray}
Taking the time average of this expression,
$\overline{\Delta E}=\lim_{t\to\infty}\frac{1}{t}\int_0^{t} \Delta E(\tau)\,d\tau$,
we obtain the following result
\begin{eqnarray}
  \overline{\Delta E}=\frac{1}{2}\,\left(1-\frac{A^2}{2}\right)\,(T-T_0)+
  \frac{A^2}{2\alpha}\,\left(\frac{\omega_*}{\omega_0}\right)^2T_0
\label{D_av}
\end{eqnarray}

For $T>T_0$ the average energy  change of the system is positive,
$\overline{\Delta E}>0$, which corresponds to  Clausius heat transfer from the  hotter
bath to the colder system.

For $T=T_0$ the average energy change of the system
is given by the second term of expression (\ref{D_av}) and  positive,
$\overline{\Delta E}>0$.
But now it corresponds to a non-Clausius heat transfer from the
bath to the system,
which occurs despite the equality of the temperatures. 

The most remarkable situation takes place when 
the system is hotter than the bath, $T_0>T$.
In that case it follows from (\ref{D_av}) that 
the direction of the average heat transfer depends on 
whether the system's initial temperature $T_0$ is higher or lower than
the characteristic temperature
\begin{eqnarray}
T_1=g_1(\alpha)\, T>T,
\end{eqnarray}
where 
\begin{eqnarray}
g_1(\alpha)=\frac{\alpha\,(2-A^2)}{\alpha (2-A^2)-2\,A^2\,(\omega_*/\omega_0)^2}=
\frac{4\alpha^3-2\alpha^2-2\alpha+1}{6\alpha^2-6\alpha+1}>1.
\label{g1}
\end{eqnarray}
Here in the second equality we used
expressions (\ref{omega*}) for the frequency $\omega_*$  and amplitude $A$ as
functions of the mass ratio $\alpha$. Except when $\alpha$ is close to one,  the factor 
$g(\alpha)$ increases approximately linearly.
For $T_0>T_1>T$, Eq. (\ref{D_av}) gives 
$\overline{\Delta E}<0$ which corresponds to 
a Clausius heat transfer: the hotter system releases heat to the  colder bath.
On the other hand, if the initial temperature of the isotope $T_0$ is 
in the interval
\begin{eqnarray}
T\le T_0<T_1
\label{condition1}
\end{eqnarray}
then the time-averaged heat transfer is non-Clausius.
In that case one finds from (\ref{D_av}) that 
$\overline{\Delta E}>0$, which suggests  that the initially 
hotter system gets heat from the  colder bath.

%The above discussion is somewhat oversimplified because  does not take into account  potential energy of the system. 
%A more detailed analysis of the next section confirms the existence of the regime of non-Clausius  heat transfer but imposes  conditions
%more restrictive than (\ref{condition1}).

%This is  in contrast with the results of the previous section
%for a uniform lattice
%and the lattice with a heavy isotope ($\alpha\le 1$).  In those  cases
%the heat transfer shows the anomalous direction only during the initial stage
%of relaxation and when averaged over a long time interval is always Clausius.

\section{Conclusion}
In this somewhat didactic paper, using the simple model of a 
harmonic chain of atoms with a 
single isotope   
and imposing a specific initial condition (of the sudden preparation type),
we demonstrate the possibility of non-Clausius heat flow
directed from the colder chain to the hotter isotope. 
The heat flow also occurs when the initial temperature of the isotope $T_0$
and that of the chain $T$ are the same.

In the chain with a heavy isotope, and in the uniform lattice,
non-Clausius heat transfer  occurs on a relatively short initial time interval
or several intervals. At longer times, the net heat transfer becomes normal,
or Clausius: it vanishes when $T=T_0$, 
and is directed from a hot subsystem to a cold one
when $T\ne T_0$.
When averaged over an asymptotically long time interval, the heat
transfer is Clausius.

In contrast, and quite remarkably, in the  chain with a light isotope a
non-Clausius heat transfer may occur  for all times  and does not vanish after
averaging over time.

For the presented model, the heat transfer properties  correlate with the
ergodic ones.
The dynamics of a tagged atom or a heavy isotope embedded in a one-dimensional
lattice is ergodic. For that case, we found that the initial non-Clausius
heat flow vanishes on long time scales. 
On the other hand, the dynamics of a light isotope in a harmonic chain is
not ergodic.
For that case, non-Clausius heat transfer does not vanish at longer times.

Our findings appear to corroborate the general  arguments of
Refs.~\cite{Gross,Dunkel,Hilbert1,Hilbert2,Campisi}:
The temperatures of two initially isolated systems $A$ and $B$ do not
completely define the dynamics of the energy flow in the composite system $AB$.
As a result,  the heat transfer between $A$ and $B$ may be non-Clausius on short
or even, in special cases, long time scales.

On the other hand, 
it remains to be seen 
to what extent our findings are generic or due to specific 
features of the exploited model. Besides being integrable, the model 
has other peculiarities which make it somewhat ambiguous to interpret 
the results in terms of thermodynamic quantities. In particular, 
in this paper we defined the direction of heat transfer based on 
the sign of the change of the system's kinetic  energy $E(t)$. 
In general, one would prefer to use the system's internal energy $U(t)$ 
instead of $E(t)$. It is natural to identify the former as the average of 
the part of the total Hamiltonian  which depends on dynamical variables 
of the system only. In our model,
that part is $H_s=P^2/(2M)+k\,Q^2$, so that  
$U(t)=\langle H_s\rangle=E(t)+k\,\langle Q^2(t)\rangle$.
At $t=0$, such a definition of the internal energy takes the reasonable 
value $U(0)=T_0$. However, for an infinite lattice the component 
$k\,\langle Q^2(t)\rangle$ can be shown to diverge in the limit $t\to\infty$ 
and thus  cannot model a physically meaningful contribution to the system's 
internal energy. The divergence of the of mean-square 
displacement $\langle Q^2(t)\rangle$ is by no means an unphysical result, 
but reflects the delocalization of a particle embedded in an 
infinite one-dimensional harmonic 
lattice~\cite{Hynes1,Weiss,Rubin,MM}. 
The divergence of $U(t)$ could be removed by including the energy 
of interaction with the bath, but in that case $U(t)$ would depend 
on bath variables as well and could hardly be qualified as an 
internal energy of the system.  One may hope to resolve 
these difficulties and limitations with models more elaborate 
than the simple one presented here.

\renewcommand{\theequation}{A\arabic{equation}}
  % redefine the command that creates the equation no.
  \setcounter{equation}{0}  % reset counter 

  \section*{APPENDIX A}  % use *-form to suppress numbering
  In this Appendix we derive the generalized Langevin equation (\ref{GLE})
  for the  isotope
for the setting depicted in Fig. 1 and described in Section II.
The first step is to diagonalize the Hamiltonian $H_b=H_r+H_l$
of the bath, Eq. (\ref{H_b}). 
Consider the Hamiltonian $H_r$ 
of the right part of the bath, Eq. (\ref{H_r}),
\begin{eqnarray}
H_r=\sum_{i=1}^N \frac{p_i^2}{2m}+
\frac{k}{2}\left\{
q_1^2+(q_2-q_1)^2+\cdots+(q_N-q_{N-1})^2+q_N^2\right\}.
\label{H_r_A}
\end{eqnarray}
It  can  be recognized as the Hamiltonian of the chain of $N+1$ atoms
with fictitious  terminal atoms $i=0$ and $i=N+1$, whose momenta and
displacements are fixed to be zero, $p_0=p_{N+1}=q_0=q_{N+1}=0$,
\begin{eqnarray}
H_r=\frac{1}{2m}\sum_{i=1}^N p_i^2+
\frac{k}{2}\sum_{i=0}^N (q_{i+1}-q_i)^2.
\end{eqnarray}
For this Hamiltonian the normal mode transformation
$\{q_i,p_i\}\leftrightarrow \{Q_j,P_j\}$ is well-known to have the form
\begin{eqnarray}
q_i=\frac{1}{\sqrt{m}}\sum_{j=1}^N A_{ij} Q_j, 
\qquad
p_i=\sqrt{m}\sum_{j=1}^N A_{ij} P_j,\qquad
i=1,2 , ... N
\label{normal_modes_transformation}
\end{eqnarray}
with the transition matrix 
\begin{eqnarray}
A_{ij}=\sqrt{\frac{2}{N+1}} \sin \frac{\pi ij}{N+1}, \qquad i,j=1,2,...N
\label{A}
\end{eqnarray}
satisfying 
the orthogonality relation $\sum_{i=1}^N A_{ij}A_{ij'}=\delta_{jj'}$.
In terms of new coordinates $\{Q_j\}$ and momenta $\{P_j\}$ the Hamiltonian
$H_r$  is diagonalized into a sum of $N$ independent normal modes with
frequencies $\omega_j$,
\begin{eqnarray}
H_r=\frac{1}{2}\sum_{j=1}^N \Big\{ P_j^2+\omega_j^2 Q_j^2\Big\}, \qquad 
\omega_j=\omega_0\,\sin \frac{\pi j}{2(N+1)},
\label{H_r2_chain}
\end{eqnarray}
where $\omega_0=2\sqrt{k/m}$.

A similar expression one finds also for the Hamiltonian $H_l$ of the
left part of the bath,
\begin{eqnarray}
H_l=\frac{1}{2}\sum_{s=1}^N \Big\{ P_s^2+\omega_s^2 Q_s^2\Big\}, \qquad 
\omega_s=\omega_0\,\sin \frac{\pi s}{2(N+1)}.
\label{H_l2_chain}
\end{eqnarray}
Here and below we use indices $1\le j\le N$ and $1\le s\le N$
to refer to normal modes for the right and left parts of the bath, respectively.

The coupling Hamiltonian $H_c$ given by Eq. (\ref{H_c}),
\begin{eqnarray}
H_c=-k\, (q_1+q_{-1})\, Q,
\label{H_c_A}
\end{eqnarray}
in terms of normal modes takes the form
\begin{eqnarray}
H_c=-\left\{\sum_{j=1}^N c_j Q_j+
\sum_{s=1}^N c_s Q_s\right\}\,Q,
\label{H_c2_chain}
\end{eqnarray}
with the coupling coefficients for the right bath
\begin{eqnarray}
  c_j=\frac{k}{\sqrt{m}}\,A_{1j}=\frac{k}{\sqrt{m}}\,\sqrt{\frac{2}{N+1}}\,
  \sin \frac{\pi j}{N+1},
\label{c_j}
\end{eqnarray}
and similar expressions for the coupling coefficients $c_s$ 
for the left bath. 

The equation of motion for the isotope
\begin{eqnarray}
\dot P=-\frac{\partial H}{\partial Q}=-2kQ+k(q_1+q_{-1})
\end{eqnarray}
in terns of normal coordinates takes the form 
\begin{eqnarray}
\dot P=-2kQ+\sum_{j=1}^N c_j Q_j+\sum_{s=1}^N c_s Q_s.
\label{eq_motion_chain}
\end{eqnarray}
Our goal is to 
find explicit expressions for normal modes $\{Q_j,Q_s\}$, to substitute them
into the above equation of motion of the isotope, and to present the latter
in the Langevin form.

The equations of motions for the normal modes of the right part of the bath read
\begin{eqnarray}
  \dot P_j=-\frac{\partial H}{\partial Q_j}=-\omega_j^2 Q_j+
  c_j\,Q,\qquad\dot Q_j=
  \frac{\partial H}{\partial P_j}=P_j.
\end{eqnarray}
Differentiating the second of these equations and substituting into the first one yields
equations for $Q_j$, 
\begin{eqnarray}
\ddot Q_j+\omega_j^2 Q_j=c_j\,Q.
\end{eqnarray}
Solving these equations using, for instance, the method of Laplace transform,
one finds
\begin{eqnarray}
  Q_j(t)=Q_j^0(t)+\frac{c_j}{\omega_j}\int_0^t\sin(\omega_j\tau)\,
  Q(t-\tau)\,d\tau.
\label{normal_modes2}
\end{eqnarray}
Here  $Q_j^0(t)$ is the general solution of the homogeneous equation 
$\ddot Q_j+\omega_j^2 Q_j=0$,
\begin{eqnarray}
Q_j^0(t)=Q_j(0)\cos(\omega_j t)+\frac{P_j(0)}{\omega_j}\sin(\omega_j t).
\label{Q_0}
\end{eqnarray}
Integrating by parts, expression (\ref{normal_modes2})
can  be written in terms of the isotope's  momentum $P=M\,\dot Q$,
\begin{eqnarray}
Q_j(t)=Q_j^0(t)+\frac{c_j}{\omega_j^2}\left\{
Q(t)-Q(0)\cos(\omega_j t)-
\frac{1}{M}\int_0^t\cos(\omega_j \tau)\, P(t-\tau)\,d\tau
\label{normal_modes3}
\right\}.
\end{eqnarray}
Substitution  of this and a similar expression for the left
normal modes $Q_s(t)$ into the impurity's equation of motion
(\ref{eq_motion_chain}),
yields the generalized Langevin equation 
\begin{eqnarray}
\dot P(t)=-\int_0^t K(t-\tau)\,P(\tau)
+\xi(t)-M\,Q(0)\,K(t)-k_*\,Q(t)
\label{GLE1_chain}
\end{eqnarray}
where the memory kernel $K(t)$ and stochastic force $\xi(t)$ are given by the expressions
\begin{eqnarray}
K(t)&=&\frac{1}M\,
\sum_{j=1}^N\left(
\frac{c_j}{\omega_j}
\right)^2\cos(\omega_j t)+
\frac{1}{M}\,
\sum_{s=1}^N\left(
\frac{c_s}{\omega_s}
\right)^2\cos(\omega_s t),
\label{kernel_A}
\\
\xi(t)&=&\sum_{j=1}^N c_j\, Q_j^0(t)+
\sum_{s=1}^N c_s \,Q_s^0(t),
\label{stochastic_force_chain}
\end{eqnarray}
with $Q_j^0(t)$ and $Q_s^0(t)$ given by (\ref{Q_0}).

Equation (\ref{GLE1_chain}) contains the initial slip
force $-M Q(0) K(t)$ and the  linear force
$-k_*Q(t)$ with the $N$-dependent 
spring constant
\begin{eqnarray}
k_*=2k-
\sum_{j=1}^N\left(
\frac{c_j}{\omega_j}
\right)^2-
\sum_{s=1}^N\left(
\frac{c_s}{\omega_s}
\right)^2.
\label{k_eff}
\end{eqnarray}
The force $-k_*Q(t)$ tends to localize the isotope about its
equilibrium position $Q=0$. 
It 
disappears in the limit of the infinite chain.
Indeed, with (\ref{H_r2_chain}), (\ref{H_l2_chain}) and
(\ref{c_j}) one finds
\begin{eqnarray}
\sum_{j=1}^N\left(
\frac{c_j}{\omega_j}\right)^2=
\sum_{s=1}^N\left(
\frac{c_s}{\omega_s}\right)^2=\frac{N}{N+1}\,k.
\label{aux1_chain}
\end{eqnarray}
Then it follows from (\ref{k_eff})  and (\ref{aux1_chain})
that $k_*\to 0$ as  $N\to\infty$, and the Langevin equation (\ref{GLE1_chain})
takes the form (\ref{GLE}), exploited  in the main text,
\begin{eqnarray}
\dot P(t)=-\int_0^t K(t-\tau)\,P(\tau)
+\xi(t)-M\,Q(0)\,K(t).
\label{GLE2_chain}
\end{eqnarray}

%For the case $m'\ge m$, 
%the disappearance of the force $-k_* q(t)$ leads to the isotope's delocalization: 
%One can show that 
%the mean-square displacement of a tagged particle
%of mass $m'$
%embedded in a one-dimensional lattice of $N$ atoms with masses
%$m\le m'$ diverges as $N\to\infty$. As a result, the isotope behaves as
%a free Brownian particle. 
%On the other hand, a light isotope, $m'<m$, remains localized even
%in the infinite chain due to the formation of a localized vibrational mode.

In order to interpret (\ref{GLE1_chain}) as a Langevin equation one
needs to define statistical properties of the stochastic force $\xi(t)$
with respect to an appropriate ensemble.  
In this paper we assume that the 
lattice at $t\le 0$ is in the state of constrained equilibrium, see Fig. 1,
with the isotope kept fixed in the equilibrium position. Respectively,
we define the average of an arbitrary dynamical variable $A$ over the initial 
coordinates and momenta of the bath 
\begin{eqnarray}
\langle A\rangle=\int
\prod_j{dQ_j dP_j}
\prod_s{dQ_s dP_s}\,(\rho_b\,A)
\label{average}
\end{eqnarray}
with the canonical distribution (\ref{rho_b}),
\begin{eqnarray}
\rho_b=Z^{-1}\,e^{-H_b/T}=
\left(Z_r^{-1}\,e^{-H_r/T}\right)\,
\left(Z_l^{-1}\,e^{-H_l/T}\right).
\label{dist1}
\end{eqnarray}
Then it is straightforward to show  that $\xi(t)$ can be interpreted
as a stochastic process which is zero-centered, stationary,
and satisfying the standard fluctuation-dissipation relation,
\begin{eqnarray}
\langle \xi(t)\rangle=0,
\qquad
\langle \xi(t)\,\xi(t')\rangle=\langle \xi(0)\,\xi(t-t')\rangle,\qquad
\langle \xi(t)\,\xi(t')\rangle=M T\, K(t-t').
\label{xi_properties}
\end{eqnarray}

As a final step, let us   derive an 
expression for the kernel $K(t)$, Eq. (\ref{kernel_A}), 
in the limit of the infinite lattice.
Taking into account the expressions for $c_j$
and $\omega_j$ one finds
\begin{eqnarray}
  \left(\frac{c_j}{\omega_j}\right)^2=\frac{2k}{N+1}\,
  \cos^2\left(\frac{\pi}{2}\frac{j}{N+1}\right)=
\frac{2k}{N+1}\,\cos^2(\varphi_j),
\label{aux3_A}
\end{eqnarray}
where we introduced the discrete variable
\begin{eqnarray}
\varphi_j=\frac{\pi}{2}\frac{j}{N+1}.
\end{eqnarray}
Substitution of (\ref{aux3_A}) into (\ref{kernel_A}) yields  
\begin{eqnarray}
K(t)=
\frac{k}{M}\,\frac{4}{N+1}\,\sum_{j=1}^N
\cos^2(\varphi_j)\,\cos\Bigl(
\omega_0 t\,\sin (\varphi_j)\Bigl).
\end{eqnarray}
With  $\Delta\varphi=\varphi_n-\varphi_{n-1}=\frac{\pi}{2}\frac{1}{N+1}$,
the above expression can be written as
\begin{eqnarray}
K(t)=
\frac{k}{M}\,\frac{8}{\pi}\,\sum_{j=1}^N
\cos^2(\varphi_j)\,\cos\Bigl(
\omega_0 t\,\sin (\varphi_j)\Bigl)\,\Delta\varphi.
\end{eqnarray}
In the limit of the infinite chain, $N\to\infty$, the above
expression takes the integral form
\begin{eqnarray}
K(t)=\alpha\,\omega_0^2\,\,\frac{2}{\pi}
\int_0^{\pi/2} \cos^2(\varphi)\,\cos(\omega_0 t\, \sin\varphi)\, d\varphi,
\end{eqnarray}
where $\alpha=m/M$.
The evaluation of the integral gives the expression
for the kernel in terms of Bessel functions
\begin{eqnarray}
K(t)=
\frac{\alpha\,\omega_0}{t}\,J_1(\omega_0 t)=\frac{\alpha\,\omega_0^2}{2}\,
\Bigl\{
J_0(\omega_0 t)+J_2(\omega_0 t)
\Bigr\}.
\end{eqnarray}
This is expression (\ref{kernel}) of the main text.

\renewcommand{\theequation}{B\arabic{equation}}
  % redefine the command that creates the equation no.
  \setcounter{equation}{0}  % reset counter 

  \section*{APPENDIX B}  % use *-form to suppress numbering
In this appendix we perform the 
inversion of 
the resolvent's Laplace transform
(\ref{resolvent_Laplace}),
\begin{eqnarray}
\tilde R(s)=\frac{1}{\alpha \sqrt{s^2+\omega_0^2}+(1-\alpha) s}
\label{resolvent_Laplace_B}
\end{eqnarray}
for arbitrary values of the mass ratio $\alpha=m/M$.
As discussed in Section IV, 
expression (\ref{resolvent_Laplace_B}) 
also gives the Laplace transform for the 
equilibrium correlation function $C(t)$ given by (\ref{C}),
which was the subject of many studies.
Although the result is well known~\cite{Rubin},
the inversion is not without subtlety and  perhaps deserves
to be discussed in a didactic manner.

The inversion of (\ref{resolvent_Laplace_B}) is given by the Bromwich integral
\begin{eqnarray}
 R(t)=\frac{1}{2\pi i} \int_{\gamma-i\infty}^{\gamma+i\infty} e^{st}\, \tilde R(s)\,ds.
 \label{Bromwich_B}
\end{eqnarray}
The function $\tilde R(s)$ has two branches, which we shall
denote as $\tilde R_1(s)$ and $\tilde R_2(s)$ and write as
\begin{eqnarray}
\tilde R_k(s)=\frac{1}{\alpha f_k(s)+(1-\alpha) s}, \qquad k=1,2
\label{branches}
\end{eqnarray}
where $f_1(s)$ and $f_2(s)$ are the two branches of the square-root function
\begin{eqnarray}
f(s)=\sqrt{s^2+\omega_0^2}=\sqrt{s+i\omega_0}\,\sqrt{s-i\omega_0}.
\end{eqnarray}
It is convenient to write in the last expression $s\pm i\omega_0$
in a polar form
\begin{eqnarray}
s-i\omega_0=r_1 e^{i\theta_1}, \qquad
s+i\omega_0=r_2 e^{i\theta_2}, 
\end{eqnarray}
with polar coordinates $(r_k,\theta_k)$ defined on the left side of  Fig. 4.
Then we can define the two branches of $f(s)$ by the expression
\begin{eqnarray}
 f_k(s)=\sqrt{r_1\,r_2}\,\, e^{i\,\frac{\theta_1+\theta_2}{2}}, \qquad k=1,2
\label{f12}
\end{eqnarray}
where the ranges of arguments $\theta_1$ and $\theta_2$
for the first branch $f_1(s)$ are the same 
\begin{eqnarray}
 -\frac{3\pi}{2}<\theta_1<\frac{\pi}{2},\qquad
-\frac{3\pi}{2}<\theta_2<\frac{\pi}{2}, 
 \label{branch1}
\end{eqnarray}
while for the second branch $f_2(s)$
\begin{eqnarray}
 -\frac{3\pi}{2}<\theta_1<\frac{\pi}{2}, \qquad
 \frac{\pi}{2}<\theta_2<\frac{5\pi}{2}.
 \label{branch2}
\end{eqnarray}
As easy to verify, the two functions $f_1(s)$ and $f_2(s)$
 defined in this way are continuous at any $s$
 except on the branch cut along the imaginary axis connecting the two branch
 points $\pm i\omega_0$.  On the branch cut both branches $f_1(s)$
 and $f_2(s)$ are discontinuous. It is instructive to identify  the following
 mapping rules for the functions $f_1(s)$ and $f_2(s)$:

 (a) Let $s=x>0$  be real and positive. Then $f_1(s)$ is also real and
 positive, while $f_2(s)$
is real and negative.

(b) Let $s=-x<0$  be real and negative. Then $f_1(s)$ is also real and negative,
while $f_2(s)$
is real and positive.

(c) Let $s=iy$, $y>\omega_0$ be on the positive part of the imaginary axis above the branch cut.
Then $f_1(s)$ and $f_2(s)$ are on 
the positive and negative parts of the imaginary axis, respectively.

(d) Let $s=-iy$, $y>\omega_0$ be on the negative part of the imaginary axis below the branch cut.
Then $f_1(s)$ and $f_2(s)$ are on 
on the negative and positive parts of the imaginary axis, respectively.

\begin{figure}[t]
\includegraphics[height=5cm]{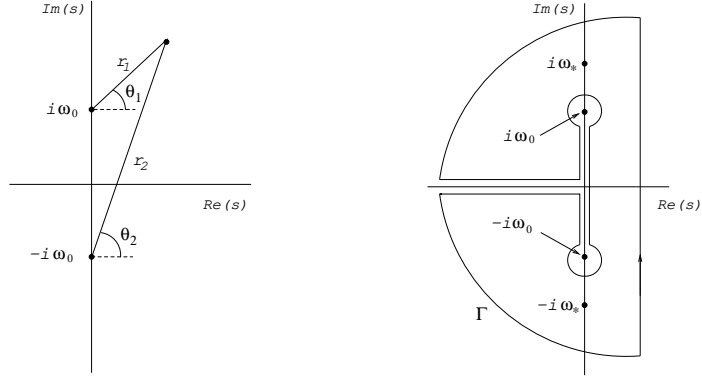}
\caption{ Left: Definition of polar coordinates $(r_1,\theta_1)$
  and $(r_2,\theta_2)$
  used in Eq. (\ref{f12}) to represent the two-value function
  $f(s)=\sqrt{s^2+\omega_0^2}$.
Right: The integration contour $\Gamma$ in the integral (\ref{I}).
The poles at $\pm i\omega_*$ exist only for the case of a light isotope,
$\alpha>1$. 
}
\end{figure}

Let us show that  only one branch 
of the function $\tilde R(s)$ is physically meaningful. Indeed,
according to (\ref{resolvent2}), $R(0)=1$. Then, using the initial
value theorem, we get the condition
\begin{eqnarray}
R(0)=\lim_{s\to \infty} s \tilde R(s)=
\lim_{s\to\infty}\frac{s}{\alpha\sqrt{s^2+\omega_0^2}+(1-\alpha) s}=1.
\label{require}
\end{eqnarray}
As follows from  the mapping rules above, 
condition (\ref{require}) is only satisfied
if the two-value function $f(s)=\sqrt{s^2+\omega_0^2}$ in the denominator
is represented by the first branch $f_1(s)$.
Therefore the Bromwich integral (\ref{Bromwich_B}) should be evaluated with 
the first branch of the function $\tilde R(s)$,
\begin{eqnarray}
  R(t)=\frac{1}{2\pi i} \int_{\gamma-i\infty}^{\gamma+i\infty} e^{st}\,
  \tilde R_1(s)\,ds,\qquad
 \tilde R_1(s)=\frac{1}{\alpha f_1(s)+(1-\alpha) s}.
 \label{Bromwich2_B}
\end{eqnarray}
Below we shall refer to the branches $\tilde R_1(s)$ and $\tilde R_2(s)$
as physical and unphysical, respectively.

For $\alpha\ne 1$, the function $\tilde R(s)$ given
by (\ref{resolvent_Laplace_B}) has two poles at positions $s_1$ and
$s_2$ which satisfy the equation
\begin{eqnarray}
\alpha \sqrt{s^2+\omega_0^2}+(1-\alpha)\,s=0.
\label{eq_for_poles}
\end{eqnarray}
The solutions
\begin{eqnarray}
s_{1,2}=\pm i\,\frac{\alpha}{\sqrt{2\alpha-1}}\, \omega_0
\label{poles}
\end{eqnarray}
are purely imaginary for $\alpha>1/2$, and real for $\alpha<1/2$.

One can show that expression (\ref{poles}) defines the poles
for the physical branch $\tilde R_1(s)$  only for $\alpha>1$,
i.e. for a light isotope, while 
for $\alpha\le 1$  expression (\ref{poles}) gives the poles
for the  unphysical branch $\tilde R_2(s)$. Indeed, let us rewrite
Eq. (\ref{eq_for_poles}) as
\begin{eqnarray}
f(s)=\sqrt{s^2+\omega_0^2}=-\frac{1-\alpha}{\alpha}\,s.
\label{aux_B1}
\end{eqnarray}
Consider first the case $\alpha>1/2$ when 
Eq. (\ref{poles}) predicts the poles located on the imaginary axis.
Consider the pole at 
$s_1=iy$, $y>0$, which is on the positive part of the imaginary axis.
As follows from (\ref{aux_B1}), $f(s_1)=\sqrt{s_1^2+\omega_0^2}$
must have a values on the  negative and positive parts of
the imaginary axis for $\alpha<1$ and $\alpha>1$, respectively.
According to mapping rule (c),  $f(s)$ must be represented
by the branch $f_2(s)$ for $\alpha<1$ and by the branch $f_1(s)$
for $\alpha>1$.  Therefore,  the pole at $s_1=iy$, $y>0$ is the one
for the unphysical branch $\tilde R_2(s)$ if $1/2<\alpha<1$ and
for the physical branch $\tilde R_1(s)$ if $\alpha>1$.

The  same conclusion we arrive at, now taking into account mapping rule (d),
for a pole at $s_2=-iy$, $y>0$ located on the negative imaginary axis.

Consider now the case $\alpha<1/2$, when 
Eq. (\ref{poles}) predicts the poles located on the real axis.
Consider the pole at $s_1=x>0$ located on the positive part of the real axis. 
As one observes from (\ref{aux_B1}), in that case
$f(s_1)=\sqrt{s_1^2+\omega_0^2}$ 
must have a value on the  negative part of the real axis.
According to mapping rule (a), this is only possible if $f(s)$
is represented by the branch $f_2(s)$. Therefore, for $\alpha<1/2$
the pole located on the positive part of the real axis is a pole
for the unphysical branch $\tilde R_2(s)$. Similarly, using mapping
rule (b), one verifies that the second pole at $s_2=-x<0$ is also
a pole for the unphysical branch $\tilde R_2(s)$.

Summarizing, for a light isotope, $\alpha>1$, the physical
branch $\tilde R_1(s)$ of the resolvent's transform  has 
two branch points $\pm i\omega_0$ and 
two poles on the imaginary axis 
\begin{eqnarray}
  s_{1,2}=\pm i\omega_*, \qquad \omega_*
  =\frac{\alpha}{\sqrt{2\alpha-1}}\,\omega_0>\omega_0.
\end{eqnarray}
For a heavy isotope, $\alpha<1$, and a uniform lattice $\alpha=1$,
the function $\tilde R_1(s)$ has two branch points $\pm i \omega_0$ and
no poles.
Since all singular points of $\tilde R_1(s)$
are on the imaginary axis, 
the integration 
in (\ref{Bromwich2_B}) is along an arbitrary vertical
line to the right of the origin.

The remaining steps are standard.
In order to evaluate the Bromwich integral (\ref{Bromwich2_B}),
consider an auxiliary integral 
\begin{eqnarray}
 I(t)=\frac{1}{2\pi i} \int_\Gamma e^{st}\, \tilde R_1(s)\,ds
 \label{I}
\end{eqnarray}
over the closed contour $\Gamma$ shown on the right in Fig. 4.
The  contributions to $I$ from the the 
large semi-arc of radius $R$ and small circles of radius $\epsilon$
about branch points $\pm i\omega_0$
vanish when $R\to\infty$ and $\epsilon\to 0$. The contribution
from the two horizontal lines above and below the real axis vanishes
as well as the distance between the lines goes to zero. The only
two non-zero contributions to $I=I_1+I_2$ come from the integration over:

1) the two sides of the vertical branch cut connecting
the branch points $\pm i\omega_0$
($I_1$),

2) the rightmost vertical segment ($I_2$). 

\noindent In the limit $R\to \infty$, the   contribution
$I_2$ equals $R(t)$, so one gets  $I=I_1+R(t)$. On the other hand,
according to Cauchy's theorem, the integral $I$ equals
to the sum of residues at the poles inside $\Gamma$,
\begin{eqnarray}
  I(t)=I_1(t)+R(t)=h(\alpha-1)\left\{{\rm Res}
  [e^{st}\tilde R_1(s), i\omega_*]+{\rm Res}
  [e^{st}\tilde R_1(s), -i\omega_*]\right\}.
\label{aux2_B}
\end{eqnarray}
Here the step function $h(x)$ is defined as
\begin{equation}
    h(x)=
    \begin{cases}
      1, & \text{if} \quad x>0 \\
      0, & \text{if} \quad x\le 0.
    \end{cases}
    \label{step}
  \end{equation}
The appearance  of $h(\alpha-1)$  in (\ref{aux2_B}) reflects
that $\tilde R_1(s)$ has the poles inside $\Gamma$ only for
a light isotope $\alpha>1$.
As discussed above, for $\alpha\le 1$ the function $\tilde R_1(s)$
has no singularities inside $\Gamma$, so that $I=0$.
From (\ref{aux2_B}) one gets,
\begin{eqnarray}
  R(t)=h(\alpha-1)\left\{{\rm Res}
  [e^{st}\tilde R_1(s), i\omega_*]+{\rm Res}
  [e^{st}\tilde R_1(s), -i\omega_*]\right\}-I_1(t).
\label{aux3_B}
\end{eqnarray}

Consider first the integral $I_1(t)$, which has two contributions,
$I_1=I_1^-+I_1^+$. The contribution $I_1^-$ is the integral
over the vertical path  just left from the branch cut, i.e.  
from $-i\omega_0-\epsilon$ to $i\omega_0-\epsilon$ with
infinitesimal $\epsilon>0$.
Using the path parametrization 
$s=iy-\epsilon$ with  $-\omega_0<y<\omega_0$, one gets
\begin{eqnarray}
I_1^-=\frac{1}{2\pi i}\int_{-\omega_0}^{\omega_0}
e^{st} \tilde R_1(s) s'(y)dy=
\frac{1}{2\pi}\int_{-\omega_0}^{\omega_0}
\frac{e^{st}}{\alpha f_1(s)+(1-\alpha)s}dy.
\end{eqnarray}
According to (\ref{f12}) and (\ref{branch1}), 
on the given path 
\begin{eqnarray}
f_1(s)=-\sqrt{r_1 r_2}=-\sqrt{(\omega_0-y)(\omega_0+y)}=-\sqrt{\omega_0^2-y^2}.
\end{eqnarray}
Then 
\begin{eqnarray}
I_1^-=
\frac{1}{2\pi}\int_{-\omega_0}^{\omega_0}
\frac{e^{iyt}\,dy}{-\alpha \sqrt{\omega_0^2-y^2}+i(1-\alpha)y}.
\label{i-}
\end{eqnarray}

Similarly, one can evaluate the contribution $I_1^+$ which
is the integral over the right side of the branch cut,
from $i\omega_0+\epsilon$ to $-i\omega_0+\epsilon$.
Now the path is parameterized  as $s=iy+\epsilon$, and 
$f_1(s)=\sqrt{r_1r_2}=\sqrt{\omega_0^2-y^2}$, which gives
\begin{eqnarray}
I_1^+=
\frac{1}{2\pi}\int_{\omega_0}^{-\omega_0}
\frac{e^{iyt}\,dy}{\alpha \sqrt{\omega_0^2-y^2}+i(1-\alpha)y}=
-\frac{1}{2\pi}\int_{-\omega_0}^{\omega_0}
\frac{e^{iyt}\,dy}{\alpha \sqrt{\omega_0^2-y^2}+i(1-\alpha)y}.
\label{i+}
\end{eqnarray}
Adding (\ref{i-}) and (\ref{i+}), and taking into account
that the contribution from the odd part of the integrand is zero, yields
\begin{eqnarray}
  I_1(t)=-\frac{2\alpha}{\pi}\int_{0}^{\omega_0}\frac{\sqrt{\omega_0^2-y^2}\,
    \cos(yt)}
{(1-2\alpha)y^2+\alpha^2\omega_0^2}\,dy.
\label{i1}
\end{eqnarray}

It remains to evaluate the residues in Eq. (\ref{aux3_B}). 
One can verify that the pole at $s_1= i \omega_*$ is  the pole
of first order, so that
\begin{eqnarray}
  {\rm Res} [e^{st}\tilde R_1(s), i\omega_*]=\lim_{s\to i\omega_*}
  e^{st}\tilde R_1(s)(s-i\omega_*)=\lim_{s\to i\omega_*}
  \frac{e^{st}(s-i\omega_*)}{\alpha\,f_1(s)+(1-\alpha)s}.
\end{eqnarray}
Evaluating the limit with  L'Hospital's rule one gets
\begin{eqnarray}
  {\rm Res} [e^{st}\tilde R_1(s), i\omega_*]=\frac{\alpha-1}{2\alpha-1}\,
  e^{i\omega_*t}.
\label{res1}
\end{eqnarray}
Similarly, for the residue at the second pole one obtains
\begin{eqnarray}
  {\rm Res} [e^{st}\tilde R_1(s), -i\omega_*]=\frac{\alpha-1}{2\alpha-1}
  \,e^{-i\omega_*t}.
\label{res2}
\end{eqnarray}
Finally, the substitution of (\ref{i1}), (\ref{res1}) and (\ref{res2})  into
(\ref{aux3_B}) yields for the resolvent the following expression
\begin{eqnarray}
R(t)=h(\alpha-1)\,A(\alpha)\,\cos(\omega_*t)+\varphi(t),
\label{result_B}
\end{eqnarray}
were the amplitude of the oscillatory term equals
\begin{eqnarray}
A(\alpha)=\frac{2\alpha-2}{2\alpha-1},
\end{eqnarray}
the function $\varphi(t)=-I_1(t)$ is given by the integral
\begin{eqnarray}
  \varphi(t)=\frac{2\alpha}{\pi}\int_{0}^{\omega_0}
  \frac{\sqrt{\omega_0^2-y^2}\,\cos(yt)}
{(1-2\alpha)y^2+\alpha^2\omega_0^2}\,dy,
\label{varphi}
\end{eqnarray}
and the step function $h(x)$ is defined by Eq. (\ref{step}).
The result (\ref{result_B}) is equivalent
to Eqs. (\ref{R_heavy}) and (\ref{sol2}) of the main text.

As discussed in Section VI, 
the function $\varphi(t)$ can be expressed in closed form in terms
of Bessel functions for $\alpha=1$  and $\alpha=1/2$. One can verify
that $\varphi(t)\to 0$ as $t\to \infty$ for any value of the mass
ratio $\alpha$.

For the purpose of the present paper, the most remarkable feature
of expression (\ref{result_B}) is that for a light isotope ($\alpha>1$)
the resolvent $R(t)$ does not vanish
at long times but oscillates with the (localized mode)
frequency $\omega_*=\alpha\omega_0/\sqrt{2\alpha-1}$. 
For  a heavy isotope or a uniform lattice ($\alpha\le 1$)
localization does not occur, and the resolvent $R(t)$ is given by
the decaying function $\varphi(t)$.

\end{document}